\documentstyle[12pt]{article}
\begin{document}
%
\newcommand{\kmug}    {\mbox{$K^+ \! \rightarrow \! \mu^+ \nu_\mu \gamma$}}
\newcommand{\sdp}    {\mbox{${\rm SD}^+$}}
\newcommand{\sdm}    {\mbox{${\rm SD}^-$}}
\newcommand{\pnn}     {\mbox{$K^+ \! \rightarrow \! \pi^+ \nu \overline{\nu}$}}
\newcommand{\pgg}     {\mbox{$K^+ \! \rightarrow \! \pi^+ \gamma \gamma$}}
\newcommand{\pmm}     {\mbox{$K^+ \! \rightarrow \! \pi^+ \mu^+ \mu^-$}}
\newcommand{\px}      {\mbox{$K^+ \! \rightarrow \! \pi^+ X^0$}}
\newcommand{\keiii}   {\mbox{$K^+ \! \rightarrow \! \pi^0 e^+ \nu_e$}}
\newcommand{\kmiii}   {\mbox{$K^+ \! \rightarrow \! \pi^0 \mu^+ \nu_e$}}
\newcommand{\pme}{\mbox{$\pi^+ \! \rightarrow \! \mu^+ \! \rightarrow \! e^+$}}
\newcommand{\vcb}     {\mbox{$V_{cb}$}}
\newcommand{\vtd}     {\mbox{$V_{td}$}}

\title{ E787: A Search for the Rare Decay \pnn } \author{ Steve
Kettell \\ {\em Brookhaven National Laboratory, Upton, NY 11786, USA}
\\ for the E787 collaboration
\footnote{see Acknowledgement} } 
\date{November 17, 1996}
\maketitle
\baselineskip=14.5pt
\begin{abstract}

Recent results from the first phase of the E787 experiment and an
update on the current status are presented. From the first phase the
limit on the \pnn\ branching ratio is BR( \pnn\ ) $ < 2.4\times10^{-9}
$. An observation of the decay \pmm\ has been made using two separate
techniques.  An observation of the decay \pgg\ has been made and the
distribution of two photon invariant mass is inconsistent with phase
space but consistent with chiral perturbation theory.

A description of recent upgrades to the detector follows. With the
upgraded detector the \pnn\ decay should soon be observable.  A
discussion of the expected sensitivity for \pnn\ during the current
running period is presented.  Possible improvements, in order to make
a significant measurement of $|\vtd |$, are discussed.  One result
from the upgraded detector is a measurement of the structure dependent
part (\sdp ) of the decay \kmug\ with a branching ratio of BR(\sdp ) =
$(1.33\pm0.12\pm0.18)\times10^{-5}$.

\end{abstract}

\baselineskip=17pt

\section{Introduction}

The rare decay \pnn\ is a flavor changing neutral current, mediated in
the standard model by heavy quark loop diagrams, and in particular, is
sensitive to the CKM matrix element $|\vtd |$.  The theoretical
prediction for this branching ratio is very clean: the hadronic matrix
element can be determined from the \keiii\ branching ratio via an
isospin transformation, the long distance contributions are
negligible, and the QCD corrections have been calculated to next to
leading log; bringing the theoretical uncertainty to 7\%\cite{the96}.

The standard model prediction for the branching ratio is
$\sim1\times10^{-10}$.  Due to the current uncertainty in \vtd\ ,
$m_t$, and \vcb\ a value for the branching ratio outside of the range
0.6---3.0$\times10^{-10}$ would be unexpected.  Over the next few
years the E787 experiment at the AGS is expected to reach a single
event sensitivity of a few$\times10^{-11}$ and should therefore
observe \pnn\ .

\section{The early experimental results}

\begin{figure}[h]
 \vspace{10.0cm}
\includegraphics{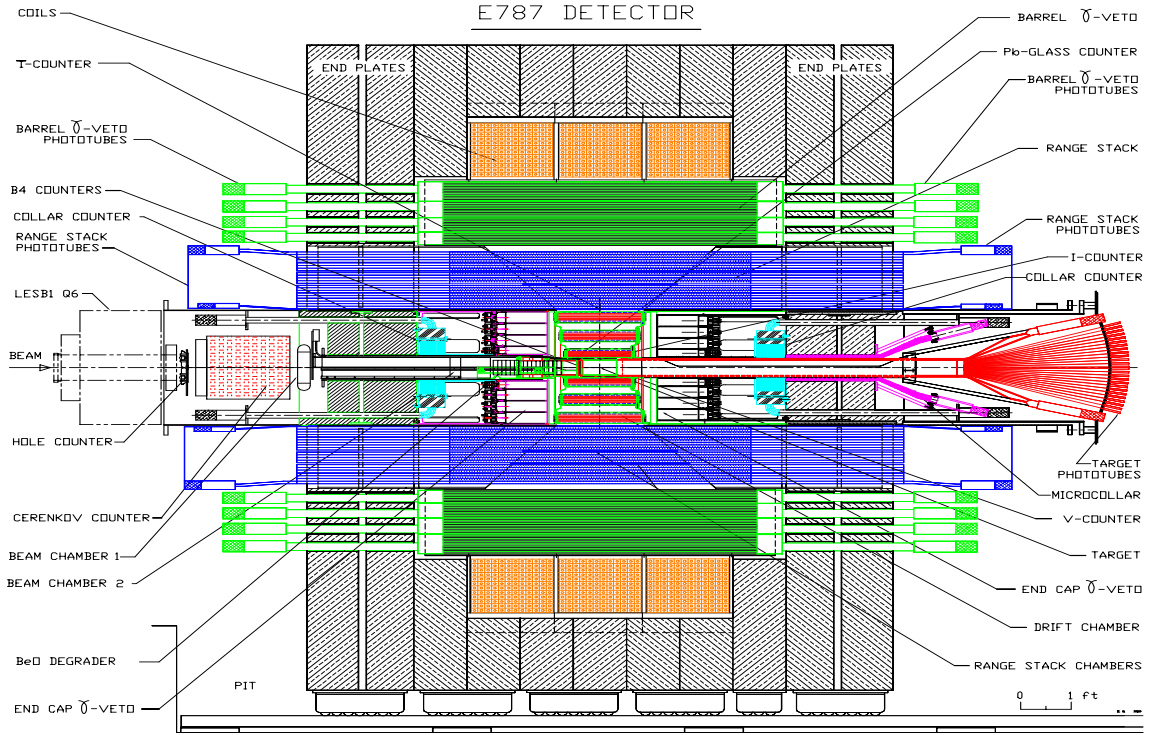}
 \caption{\it
      The E787 Detector.
    \label{det} }
\end{figure}

The early version of the detector is described elsewhere\cite{det92},
but a schematic of the upgraded version shows essentially the same
functionality (see Fig.\ref{det}).
An 800 MeV/c K$^{+}$ beam, incident from the left, is stopped in a
scintillating fiber target in the center of the 10kG solenoidal magnet
field.

The experimental signature for \pnn\ is a single charged kaon track
incident and a single charged pion track outgoing, with two missing
neutrinos. Given the weak constraints and the small branching ratio,
the particle identification and kinematics of the $\pi$ must be very
well measured and any additional particles must be vetoed with high
efficiency.  This is most readily solved in the center of mass frame;
therefore, E787 runs in a stopped kaon beam at the AGS.  The two
dominant decay modes, K$_{\pi2}$ and K$_{\mu2}$, produce
mono-energetic charged particles which can be kinematically separated
from \pnn\ . The E787 detector uses redundant measures of the
kinematics: momentum, energy and range.  The K$_{\pi2}$ can also be
suppressed by vetoing on the $\pi^0$ photons, so the detector is
surrounded by a nearly 4$\pi$ photon veto (PV) and in fact almost the
entire detector not traversed by the $\pi^{+}$ is used as a veto. The
K$_{\mu2}$ can be suppressed by requiring that the $\pi$ decay to a
$\mu$ in the range stack (RS). The entire \pme\ decay chain is
observed with 500 MHz, 8-bit transient digitizers (TD) sampling the
output of the RS scintillators.  The next significant background comes
from scattered beam pions. The primary tools for rejecting these
events are fast, finely segmented beam counters and good tracking in
the fiber target.

\subsection{ Search for \pnn }

\begin{figure}[h]
 \vspace{8.cm}
\includegraphics{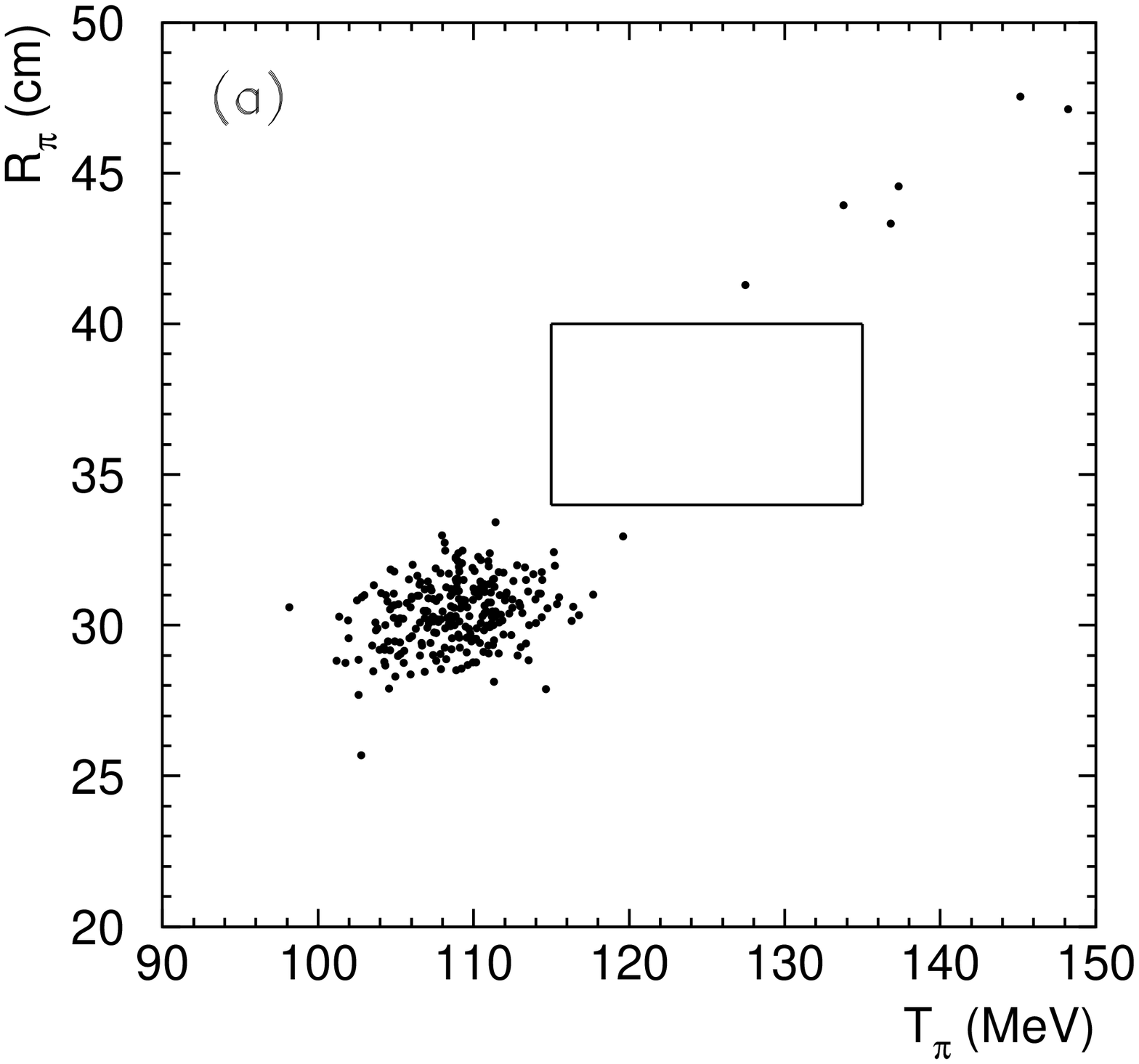}
 \caption{\it
      Distribution of \pnn\ candidates after all cuts.
    \label{pnn_fig} }
\end{figure}

A significant \pnn\ data set was collected with the original detector
during the period from 1989--1991.  The analysis of the this data is
described in detail elsewhere.\cite{pnn96} A total of
3.49$\times10^{11}$ kaons were stopped (KB$_L$).  A final plot of the
range and kinetic energy of events remaining after all cuts is shown
in Fig.\ref{pnn_fig}.
No events were seen, implying a 90\% CL limit on the branching ratio
of:
\[ BR( \pnn ) < 2.4 \times 10^{-9}\]
\[BR( \px ) < 5.2 \times 10^{-10}\]

\subsection{ Observation of the decay \pmm }

\begin{figure}[h]
 \vspace{7.25cm}
\includegraphics{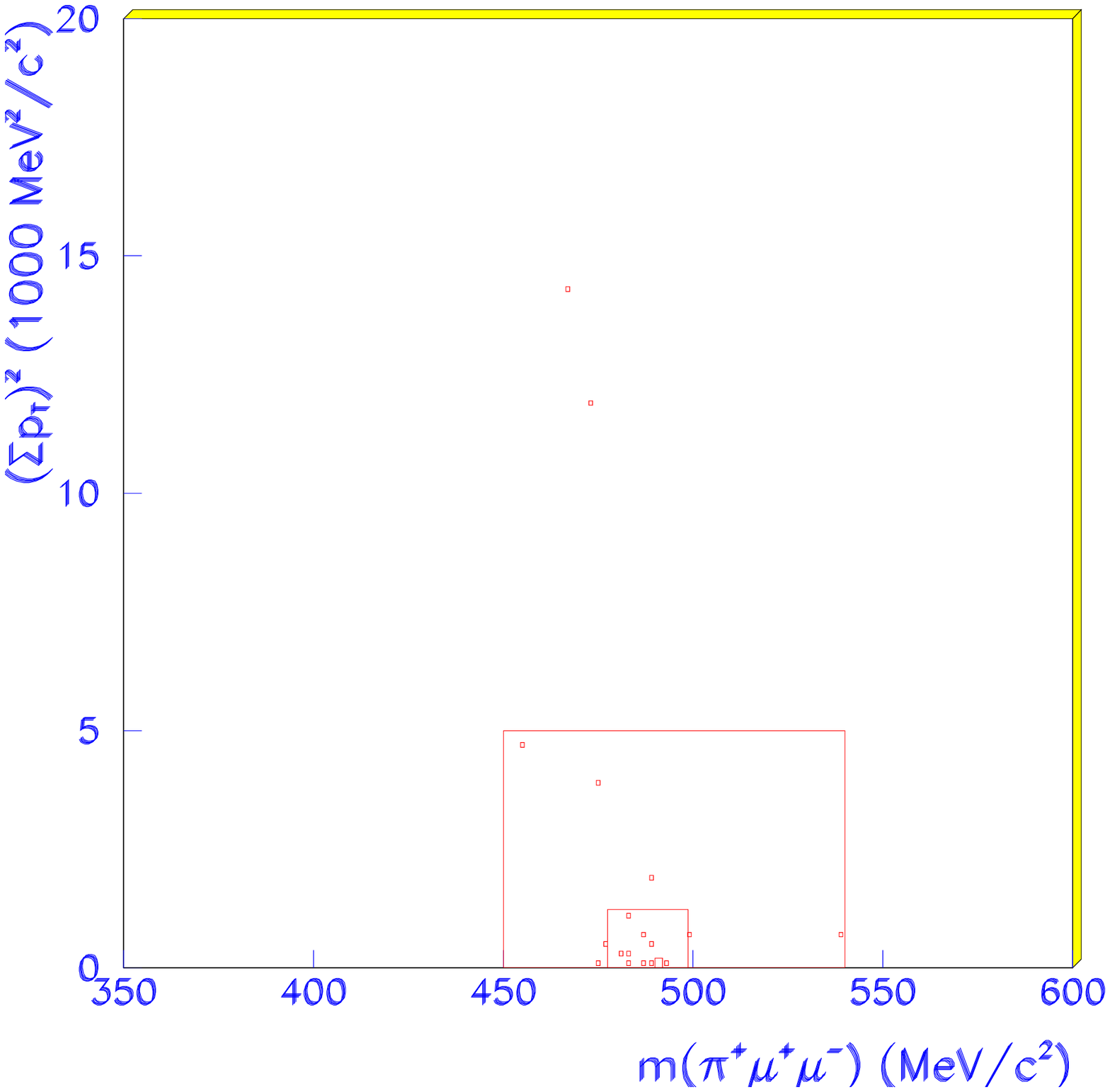}
\includegraphics{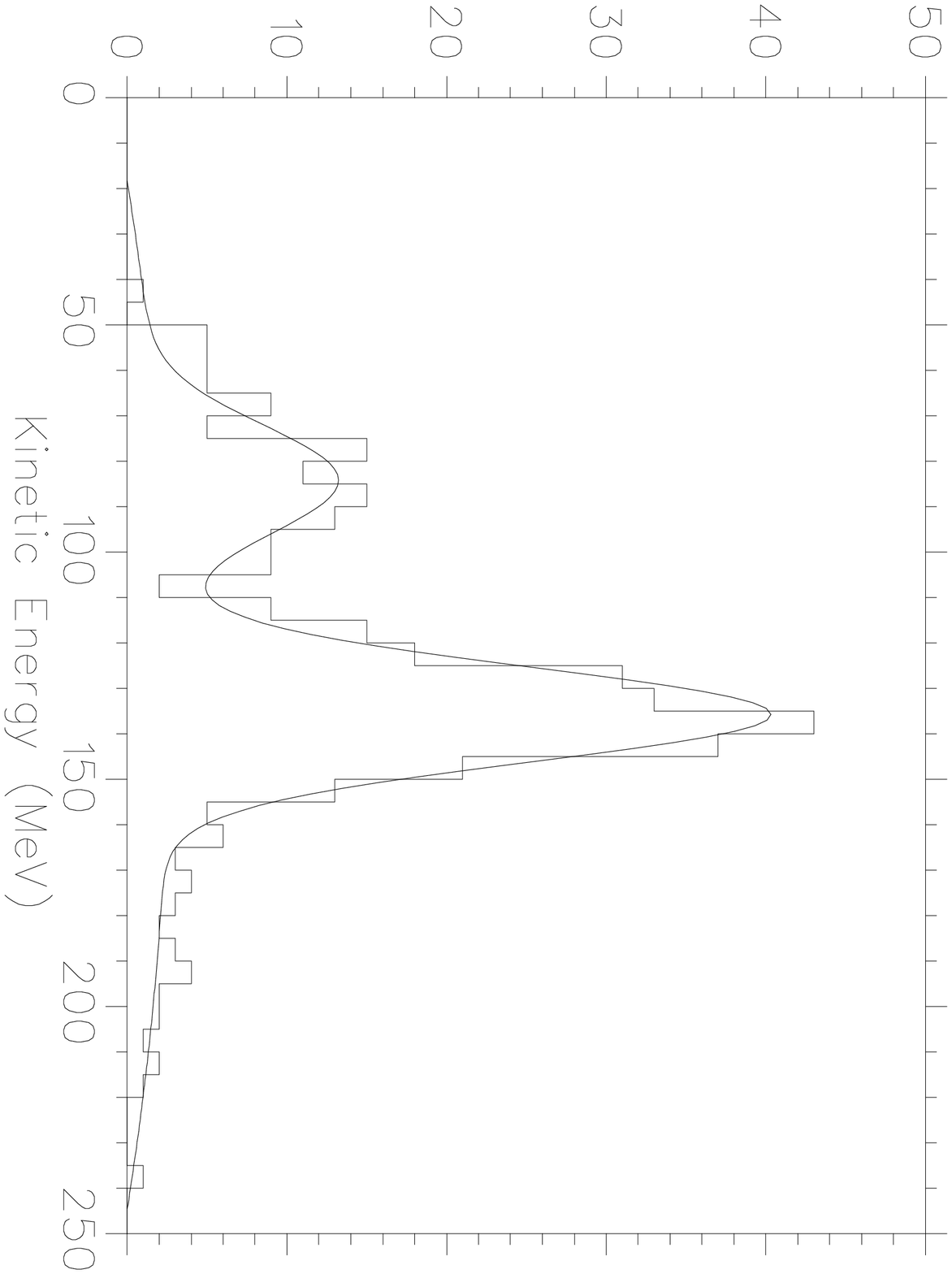}
 \caption{\it
      Final plots for \pmm\ .
    \label{pmm_fig} }
\end{figure}

A sample of $6\times10^6$ \pmm\ candidates was collected during the
1989--91 running period (KB$_L = 3.03\times10^{11}$).\cite{pmm95} The
\pmm\ trigger required two charged tracks to reach the range stack,
but not to penetrate very deeply, and required that there be no
activity in the PV. A total of $6\times10^6$ triggers was collected
during 1989--91.  Two separate techniques of analyzing the data were
employed.  One analysis required that all three tracks be
reconstructed in the drift chamber.  A second analysis, with larger
acceptance, did not require that the 3$^{rd}$ be fully
reconstructed. If the 3$^{rd}$ track remains in the target no momentum
information is available, but the total kinetic energy (Q)
is. Figure~\ref{pmm_fig} shows the final samples.
The left figure shows the events that survive all cuts of the first
 analysis as a scatter plot of the reconstructed kaon mass vs. the
 square of the transverse momentum.  A total of 13 events were seen in
 the small signal box, at low p$_T^2$ and M$_K$.  The larger box
 (excluding the smaller one) is used for a background estimation. It
 is dominated by K$_{e4}$ that have passed the 90\% confidence level
 electron rejection cut.  The background estimate is $2.5\pm1.1$
 events in the signal box.  The plot on the right shows the kinetic
 energy of the final sample of events from the second analysis.  In
 this case we see a clear peak at the value of Q = 143 MeV from the
 \pmm\ decay, as well as a smaller peak at Q = 75 from the K$_{\pi3}$
 decay.  A fit to the signal plus background yields $196.0\pm16.7$
 signal events.

\subsection{ Observation of the \pgg\ decay }

\begin{figure}[h]
 \vspace{8.0cm}
\includegraphics{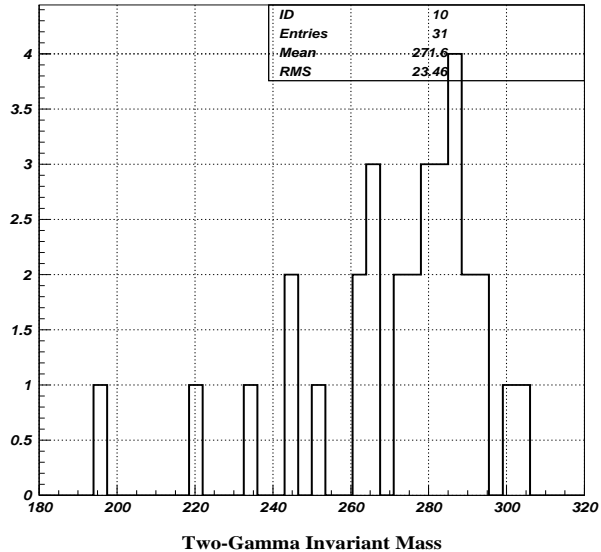}
 \caption{\it
      Fitted $M_{\gamma\gamma}$ of events surviving cuts.
    \label{pgg_fig} }
\end{figure}

A total of $2.7\times10^{6}$ \pgg\ triggers were collected (KB$_L =
1.01\times10^{10}$) during the 1991 run.\cite{pgg96} The trigger
selected $\pi^{+}$ with short range and di-photons with an opening
angle $> \sim 90^{\circ}$ and an invariant mass between 190--600
MeV/c$^2$.  The major backgrounds are from two classes. The first is
K$_{\pi3}$ and K$_{\pi2\gamma}$ with overlapping photons.  These are
rejected by cuts targeting the $\phi$ and $z$ consistency of the
photon clusters. The second class includes K$_{\mu3}$, K$_{\pi2}$,
K$_{\pi3}$ and K$_{\pi2\gamma}$; these are all kinematically distinct
from \pgg\ . The K$_{\mu3}$'s are missing a neutrino and the $\mu$
must be misidentified as a $\pi$. The $\pi$ in K$_{\pi2}$ must scatter
in the target and be down-shifted in momentum. The K$_{\pi3}$ and
K$_{\pi2\gamma}$ are missing a photon and do not conserve
momentum. After all cuts targeting these backgrounds are applied, 31
events survive. After a background subtraction, the signal is
$26.4\pm6.2$ events.  A plot of $M_{\gamma\gamma}$ of the remaining
events is shown in Fig.\ref{pgg_fig} The distribution is significantly
different than phase space and is consistent with Chiral Perturbation
Theory ($\chi$PT).

A second sample of \pgg\ with a long range pion and a low
M$_{\gamma\gamma}$ found no events and set a limit of BR(\pgg\ ,
$|p_\pi|<217$ MeV/c) $< 5.1\times10^{-7}$, assuming a phase space
distribution.  A small branching ratio for $|p_\pi|<217$ MeV/c is
consistent with $\chi$PT.

\section{The Current Status of E787}

Following the 1991 run a number of upgrades to the E787 experiment
were undertaken.  With the increased proton intensity available from
the AGS with the Booster, the beam line and essentially every detector
system were upgraded to handle higher rates and to maintain and
improve the background rejection. A drawing of the upgraded E787
detector is shown in Fig.\ref{det}

\subsection{Upgrades}

The first of these was the construction of a new low energy separated
K$^{+}$ beam-line (LESB3).\cite{les96} The central drift chamber was
replaced with a new ultra-thin chamber (UTC).  The momentum resolution
for K$_{\pi2}$ and K$_{\mu2}$ of 1.2\% and the $z$ resolution of 1 mm
are both improved by $\times$ 2 compared to the previous drift
chamber.\cite{utc96} The Pb-scintillator endcaps (EC) were replaced
with pure CsI crystals. The timing resolution is improved by $\times$
2--3; with $\sigma_T \sim$ 600ps for $\gamma$'s with E $>$ 20
MeV.\cite{csi95} The EC was moved closed to the UTC to increase the
coverage at the thinner corners at $\sim 45^{\circ}$.  Additional
Pb-scintillator photon veto systems were installed behind the endcap:
the collar (CO) and micro-collar (CM).  The RS signals were more
finely segmented, with a PMT on each end of each scintillator. This
provides better dE/dx resolution and more light ($\times$1.3 p.e.).
The tracking chambers within the RS were replaced with very low mass
straw chambers (RSSC). The target was replaced with a much brighter
scintillating fiber target (TT). Tests of the new fibers show $\times$
4--5 more light ( 20 p.e./MeV , $L_{atten}$ = 2 m ).  An additional
beam chamber was added, (BWPC2). A new \v{C}erenkov (C$_{K}$ and
C$_{\pi}$) counter with more light collection was installed.  A new
transient digitizing system based on GAs CCD's sampling with 8 bits at
500 MHz was instrumented on the EC, TT and beam elements with ranges
of 512 ns, 256ns, 256 ns respectively.\cite{ccd96} A new Fastbus based
trigger, including a fast L1.1 trigger, was built. The L1.1 is based
in an ASIC on the TD\cite{td} boards and makes a pulse area to pulse
height comparison to reject 90-95\% of $\mu^+$ in 10-20 $\mu$sec.  The
data acquisition system was upgraded to RISC based
processors\cite{daq} (12 CPU SGI Challenge) with a specially built VME
to Fastbus interface\cite{vme}. The system is consistently able to
sustain an upload of 24 Mbytes in the 2 seconds between spills.

\subsection{ \kmug }

The internal bremsstrahlung component (IB) of \kmug\ has been well
measured.\cite{aki} However, the structure dependent part, \sdp\ and
\sdm\ , ($\gamma$ radiated from internal hadron lines) had never been
observed.  A two day period of data taking in 1994 ( KB$_L =
9.2\times10^{9}$ ) was dedicated to a measurement of \sdp\ , which has
an energetic $\mu$ and an energetic $\gamma$.\cite{kmg96}

The primary backgrounds are K$_{\pi2}$ and K$_{\mu3}$ with a missed
$\gamma$ and K$_{\mu2}$ with an accidental $\gamma$. These are
suppressed by requiring the $\mu^{+}$ momentum to be above the \kmiii\
endpoint of 215 MeV/c, by vetoing on additional photons, and by
requiring that the kinematics are consistent with \kmug\ . After all
cuts the background was reduced to $105\pm6$ events out of 2693 in the
signal region. A fit to the background, IB, \sdp\, \sdm\ and
interference components yields a branching ratio:
\[ BR(SD^{+}) = ( 1.33 \pm 0.12 \pm 0.18 ) \times 10^{-5} \]
which gives a determination of the form factor $ | F_{V} + F_{A} | =
0.165 \pm 0.007 \pm 0.011 $.  The O(p$^4$) $\chi$PT prediction for the
branching ratio is $ BR(SD^{+}) = 9.2 \times 10^{-6} $.\cite{bij}

As a check, a measurement of the Inner Bremsstrahlung component of
\kmug\ gives $BR(IB) = (3.57\pm0.27) \times 10^{-4}$, which can be
compared to the theoretical value $BR(IB, E_{\mu}>100 MeV, E_{\gamma}>
20 MeV) = 3.30\times10^{-4}$.

\subsection{Expected Sensitivity for \pnn\ }

A commissioning run with most of the new detector in 1994 produced new
results on \kmug\ and provided a shakedown of most of the new
detector.  The two modestly long running periods in 1995 and 1996 ( 24
and 17 weeks) have provided a significant \pnn\ data sample. The
accumulated number of stopped kaons were $1.55\times10^{12}$ and
$1.15\times10^{12}$ respectively.  Preliminary studies indicate that
the backgrounds can be kept to the $10^{-11}$ level with acceptances
comparable to the previous years.

\section{Future Plans}

During the recent AGS2000 workshop a plan for making a 20\%
measurement of the \pnn\ branching ratio (10\% measurement of $|\vtd
|$) was outlined.\cite{ags96} The plan envisioned a factor of 10
increase in sensitivity per year (allowing an observation of several
events per year). In order to make solid extrapolations of acceptance
and background rejection, no significant increases in rates are
planned.  The major improvement factors come from reducing the beam
momentum (and increasing the fraction of usable kaons that stop in the
target) and from increasing the duty factor of the AGS. Such a plan,
with modest upgrades to the E787 experiment would allow the CKM model
of CP violation to be critically tested.

\begin{figure}[h]
 \vspace{8.8cm}
\includegraphics{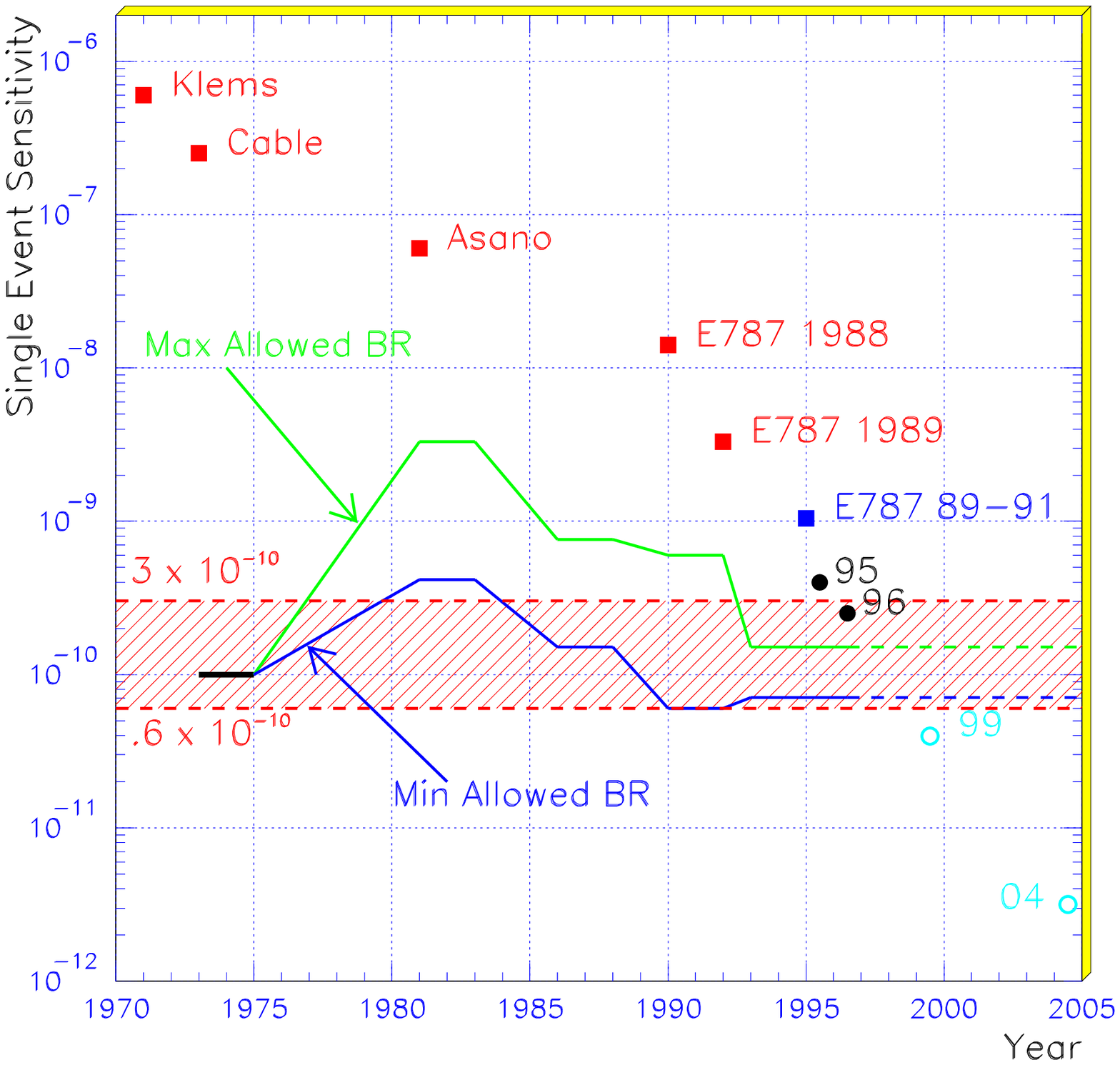}
 \caption{\it
        Past and Future Progress in searching for \pnn\ .  
    \label{br_time} }
\end{figure}

A history of the limits on the \pnn\ branching ratio is shown in
Fig.\ref{br_time}
for measurements through 1991, with some extrapolations to the near
future, and some estimates of what could be done in the AGS2000
era. The curves show the evolution of the theoretical predictions over
time.

\section{Conclusions}

The E787 experiment has reached a very exciting time, as the
sensitivity of the experiment is now very close to the standard model
prediction for \pnn\ and the experiment should expect to see events
soon.

\section{Acknowledgment}

The E787 collaboration, which is responsible for the work presented
here:\\ S.~Adler, M.S.~Atiya, I-H.~Chiang, M.~Diwan, J.S.~Frank,
J.S.~Haggerty, S.H.~Kettell, T.F.~Kycia, K.K.~Li, L.S.~Littenberg,
C.~Ng, A.~Sambamurti, A.J.~Stevens, R.C.~Strand and C.H.~Witzig: {\em
Brookhaven National Laboratory} \\ M.~Kazumori, T.~Komatsubara,
M.~Kuriki, N.~Muramatsu, H.~Okuno, A.~Otomo, S.~Sugimoto, and K.~Ukai:
{\em INS, University of Tokyo} \\ M.~Aoki, T.~Inagaki, S.~Kabe,
M.~Kobayashi, Y.~Kuno, T.~Sato, T.~Shinkawa and Y.~Yoshimura: {\em
KEK, National Laboratory for High Energy Physics} \\ W.C.~Louis: {\em
Los Alamos National Laboratory}\\ Y.~Kishi, T.~Nakano and T.~Sasaki:
{\em Osaka University} \\ D.S.~Akerib, M.~Ardebili, M.R.~Convery,
M.M.~Ito, D.R.~Marlow, R.A.~M\raisebox{.5ex}{c}Pherson, P.D.~Meyers,
M.A.~Selen, F.C.~Shoemaker, A.J.S.~Smith and J.R.~Stone: {\em
Princeton University} \\ P.~Bergbusch, E.W.~Blackmore, D.A.~Bryman,
L.~Felawka, P.~Kitching, A.~Konaka, J.A.~Macdonald, J.~Mildenberger,
T.~Numao, P.~Padley, J-M.~Poutissou, R.~Poutissou, G.~Redlinger,
J.~Roy, M. Rozon, R.~Soluk and A.S.~Turcot: {\em TRIUMF}

This work was supported by the U.S. Department of Energy under
contract no. DE-AC02-76CH00016.


\end{document}